\documentclass[aip,jcp,amsmath,preprint,reprint]{revtex4-1}
\usepackage{amsmath}
\usepackage{dcolumn}
\usepackage{hyperref}
\usepackage{graphicx}

\renewcommand{\vec}[1]{{\rm\bf #1}}
\newcommand{\im}{{\rm i}}
\newcommand{\de}{\,\mathrm{d}}
\newcommand{\al}{\alpha}
\newcommand{\ka}{\kappa}
\newcommand{\la}{\lambda}
\newcommand{\om}{\omega}
\newcommand{\ep}{\epsilon}
\newcommand{\ee}{\varepsilon}
\newcommand{\EE}{\mathcal{E}}
\newcommand{\el}{\mathcal{L}}
\newcommand{\si}{\sigma}
\newcommand{\pfi}{\varphi}
\newcommand{\EM}{\mathcal{M}}
\newcommand{\EN}{\mathcal{N}}

\begin{document}

\title{Spherical-harmonic-analysis-based optimization of atomic weighting functions \\
for multicenter numerical integration in molecules}

\author{Dimitri N. Laikov}
\email[]{laikov@rad.chem.msu.ru}
\homepage[]{http://rad.chem.msu.ru/~laikov/}
\affiliation{Chemistry Department, Moscow State University,
119991 Moscow, Russia}


\begin{abstract}
The well-known spatial integration schemes in molecular electronic structure theory,
immune to cusps and point singularities of some kind at atomic positions,
use a set of weighting functions
to split the integrand into a sum of atom-centered parts,
each dealt with in its own spherical coordinate system.
Here, for a given set of integrands in the two-center case,
a quality measure of the weighting functions
is defined to compare, design, and optimize them,
it is roughly proportional to the average number of angular quadrature points
needed to reach a given integration accuracy.
A study of Becke's fuzzy Vorono\"i cells
has helped to improve their performance by a new modification.
New spherically-symmetric unnormalized weighting functions are found
in the form of a negative power
times the negative exponential of the fourth power
of the scaled distance to the atomic center,
with the parameters related to the asymptotic decay of the integrand
and the integration accuracy --- these are much simpler
but no less efficient and naturally fit
for linear-scaling calculations.
Radial distribution of spherical quadrature orders is studied.
A radial integration scheme of double exponential type is optimized.
A symmetric analog of the pseudospectral approximation is used
for the seminumerical evaluation of two-electron repulsion integrals.
Taken together, this allows efficient calculation
of all molecular integrals with well-controlled accuracy,
as shown by tests on a set of molecules.
\end{abstract}

\maketitle

\section{Introduction}

Three-dimensional numerical integration is a helpful tool
for molecular electronic structure calculations:
on one hand, it is the best and almost the only way
to deal with the exchange-correlation models
of density functional~\cite{KS65} theory;
on the other hand, even when the analytic solutions are known,
it can greatly speed up the evaluation of the direct and exchange
two-electron terms~\cite{F87,F88,MC93,MBFR95,IN11} of wavefunction methods,
and even the many-electron integrals~\cite{BBT09,BT12}
of explicitly correlated approaches;
furthermore, it is higly vectorizable and parallelizable.
The multicenter nature of the integrand,
that may have cusps or point singularities at each atomic center,
makes the design of a good cubature rule for polyatomic molecules
more than a worthwhile mathematical exercise.
Two main paths have been followed:
a division of space into atomic spheres and interstitial regions~\cite{BVB88}, without overlap,
each with its own grid of points; or the use of atomic weighting functions
to split~\cite{BR66,B88} the integrand into a sum of well-behaved overlapping atom-centered functions,
each of which is dealt with in its own spherical coordinate system.
It is the latter that we study here,
the main idea was born~\cite{BR66} when the computational chemistry was in its childhood,
but 22 years later it began to grow into a heavy-load workhorse
after the fuzzy Vorono\"i~\cite{V08} cells were used
to build~\cite{B88} the now-standard weighting functions ---
despite their formal cubic scaling,
they were quickly adapted~\cite{SSF96} for linear-scaling calculations,
a more detailed study~\cite{LKO18} has later shown how to overcome their limitations
more carefully with accuracy in mind.
For periodic systems,
spherical (unnormalized) weighting functions in the form
of inverse third power times negative exponential of the distance
have been reported~\cite{FPV13} to work,
and yet another form~\cite{L97} of this kind has long been used for isolated systems.

It is natural to ask whether there is some best form
of the atomic weighting functions
and how to find it.
Here, we find an answer by understanding
that it is the angular integration that determines the performance ---
with an adaptive~\cite{AW92} choice of the order of quadratures~\cite{L76,LL99}
for each spherical shell to meet a given integration accuracy,
it is the number of angular points that grows the fastest 
and is most sensitive to the integrand's behavior.
As there is no well-defined orientation of the spherical grids
that could always smoothly follow the changes in molecular geometry,
a rather high accuracy may often be needed
to get rotationally-invariant molecular energies and their derivatives
free of random noise.
We take the two-center (diatomic) case as a model,
define a measure of the weighting functions' fitness,
and use it to design and optimize them
--- a sound scheme should then work
in the general polyatomic case as well
and can be tested on realistic systems.

Beside the weighting functions,
a fully-fledged molecular integration scheme also needs
a radial distribution of spherical quadrature orders around each atom
and a good radial quadrature itself.
We find an analytic fit to the distribution of orders
in the diatomic case and use it to make
a geometry-dependent polyatomic generalization
that is more economical than the traditional pruning~\cite{GJP93}.
A number of radial grids~\cite{B88,TA95,MN96,LMG01} are in widespread use,
the idea of double-exponential~\cite{TM73} integration
has also found its way~\cite{M11} into this field ---
here, we have further optimized one such scheme
and derived the estimates of its accuracy and convergence.

\section{Methodology}

Let $\{\vec{R}_i\}$ be the atomic positions in a molecule,
but we will first study a \textit{diatomic} fragment
with its cylindrically-symmetric weighting function
\begin{equation}
w_{ij}(\vec{r}) =
w\bigl(\left|\vec{r} - \vec{R}_i \right|, \left|\vec{r} - \vec{R}_j \right|, \left|\vec{R}_i - \vec{R}_j \right| \bigr)
\end{equation}
as a starting for polyatomic generalizations.
The function~$w(r_i, r_j, R) \ge 0$ of the three distances
should have the properties
\begin{eqnarray}
w(r_i, r_j, R) + w(r_j, r_i, R) &=& 1, \\
w(0,R,R) &=& 1, \\
w(R,0,R) &=& 0,
\end{eqnarray}
and its best form is to be found.
A good $w(r_i, r_j, R)$ should be localized around $r_i = 0$
and suppress singularities of the integrands
as $r_j \to 0$ of the kind up to $r_j^{-2}$,
and the smoother the better.

We make a set of $2N$ model test functions
\begin{equation}
\label{eq:fm}
f_n(r_i, r_j, R) =
\left\{
\begin{array}{lc}
w(r_i, r_j, R) b_n(r_i), & 0 \le n < N, \\
w(r_i, r_j, R) b_{n-N}(r_j), & N \le n < 2N, \\
\end{array}
\right.
\end{equation}
with the normalized radial parts
\begin{equation}
\label{eq:fa}
b_n(r) = \frac{\al_n^3}{\pi^{3/2}} \exp\left(-\al_n^2 r^2 \right)
\end{equation}
and densely-spaced even-tempered exponents
\begin{equation}
\label{eq:an}
\al_n = \al\cdot 2^{n/M} ,
\end{equation}
this is a fairly good model of a whole set of two-center molecular integrands,
simplified by dropping the low-order polynomial factors
that are too well-behaved to play a big role in what follows.
We take $M=16$ in Eq.~(\ref{eq:an}) which is more than enough, $M=8$ is almost as good.
In all cases studied, $\al_n > 2^6$ make zero contribution,
so we work with the sets $\al \le \al_n \le 2^6$
having only one parameter $\al$.

As the weighting function $w(r_i, r_j, R)$
will be optimized for the integration in spherical coordinates centered at $r_i = 0$,
we get first the coefficients from the spherical harmonic analysis
\begin{equation}
\label{eq:aln}
A_{ln} (r, R) =
\int\limits_{-1}^1 \bar{P}_l(z) f_n\left(r,\sqrt{r^2 - 2Rrz + R^2},R \right)  \de z ,
\end{equation}
made simple thanks to the cylindrical symmetry of the problem,
using the normalized Legendre polynomials
\begin{eqnarray}
\bar{P}_l(z) &=& \sqrt{l+\tfrac12\,}\, P_l(z), \\
P_0(z) &=& 1, \\
P_1(z) &=& z, \\
l P_l(z) &=& (2l - 1) z P_{l-1} (z) - (l - 1) P_{l-2} (z) .
\end{eqnarray}
A Gau{\ss}--Legendre quadrature of a high enough order
can be used to integrate numerically over $z$ in Eq.~(\ref{eq:aln}),
but the highest precision and fastest convergence can be reached
by a double-exponential~\cite{TM73} integration,
we change the variable
\begin{equation}
\label{eq:zx}
z = \tanh\left(\frac{2\sqrt{3}}{\sqrt{4 - \pi}}\cdot\frac{x}{1 - x^2}\right)
\end{equation}
and apply the trapezoidal rule for $-1<x<1$.

A (hopefully exponential) convergence of $|A_{ln}|$ as $l\to\infty$
can be quantified by the residuals
\begin{equation}
\label{eq:qln}
Q_{Ln} (r, R) = 2\pi r^3 \cdot
\left(\sum\limits_{l=L}^\infty A^2_{ln} (r, R) \right)^{1/2},
\end{equation}
that can also be computed as
\begin{equation}
\label{eq:qln2}
Q_{Ln} (r, R) = 2\pi r^3 \cdot
\left(B_n (r, R) - \sum\limits_{l=0}^{L-1} A^2_{ln} (r, R) \right)^{1/2},
\end{equation}
\begin{equation}
\label{eq:bn}
B_n (r, R) =
\int\limits_{-1}^1 f^2_n\left(r,\sqrt{r^2 - 2Rrz + R^2},R \right)  \de z .
\end{equation}
For lower-precision work, we use the more stable Eq.~(\ref{eq:qln})
together with the Gau{\ss}--Legendre rule for Eq.~(\ref{eq:aln});
for arbitrary precision computation, however,
we use Eqs.~(\ref{eq:qln2}) and~(\ref{eq:bn})
together with the double-exponential integration
over $z$ through Eq.~(\ref{eq:zx})
in both Eqs.~(\ref{eq:aln}) and~(\ref{eq:bn}).

The greatest value over the test functions
\begin{equation}
\label{eq:ql}
Q_L (r, R) = \max\limits_n Q_{Ln} (r, R)
\end{equation}
is a measure of how well the weighting functions work.
Instead of picking up the greates value from the set of $2N$,
however dense it may be, a full maximization of $Q_{Ln} (r, R)$
with respect to $\al_n \ge\al$ in Eq.~(\ref{eq:fa}) can be done numerically
to reach the limit of Eq.~(\ref{eq:ql}) as $N,M\to\infty$,
we do so when we need the highest prcision.

To simplify the optimization, a continuous function is made
from the discrete values of $Q_L (r, R)$
through the piecewise linear interpolation
\begin{eqnarray}
\label{eq:q}
\nonumber
Q(L, r, R) &=&
 \big( L - \lfloor L\rfloor\big)
 \left( Q_{\lfloor L + 1\rfloor} (r, R) -
        Q_{\lfloor L\rfloor} (r, R) \right)
\\
&+&
Q_{\lfloor L\rfloor} (r, R) ,
\end{eqnarray}
the inverse function $\el(r, R, \ee) \ge 0$ can then be found from
\begin{equation}
\label{eq:qe}
Q\big(\el(r, R, \ee), r, R\big) = \ee ,
\end{equation}
meaning the order of angular quadrature needed to integrate all test functions
to within a given error tolerance $\ee$.
We define our objective function
\begin{equation}
\label{eq:nre}
\EN (R, \ee) =
\sum\limits_{k=-\infty}^{+\infty} \EN (r_k, R, \ee)\, d_k,
\end{equation}
\begin{equation}
\label{eq:rre}
\EN (r, R, \ee) =
\big[\max\big(\el(r_k, R, \ee), l \big) + 1 \big]^2 - (l + 1)^2 ,
\end{equation}
as roughly proportional to the number of angular integration points
(beyond order $l$, we set $l=1$)
on all spherical shells for $0\le r < \infty$,
and a further average
\begin{equation}
\label{eq:ne}
\EM (\ee) =
\sum\limits_{k=0}^\infty \EN (R_k, \ee) D_k
\end{equation}
over all $R_0 \le R < \infty$,
with a simple discretization
\begin{equation}
\label{eq:k}
r_k = 2^{(k + t)/K}, \qquad
d_k = (\ln 2)/K
\end{equation}
(we set $t=0$ for now, but will vary it later),
\begin{eqnarray}
\label{eq:rk}
R_k = R_0 \cdot 2^{k/K}, \qquad
D_k = R_0 \cdot (\ln 2)/K ,
\end{eqnarray}
and a natural $R_0 = 1\mbox{ au}$.
The sum in Eq.~(\ref{eq:nre}) is only formally infinite since
for both $k \ll -K$ and $k \gg K$ all terms go quickly to zero
(with $l=1$ in Eq.~(\ref{eq:rre}), but with $l=0$ the convergence
would have been too slow),
and the same is true for Eq.~(\ref{eq:ne}).
We take $K=16$ in both Eqs.~(\ref{eq:k}) and~(\ref{eq:rk})
that is enough to integrate the functions to about 36 bits of precision.

Both $\EN (R, \ee)$ of Eq.~(\ref{eq:nre}) and $\EM (\ee)$ of Eq.~(\ref{eq:ne}) are functionals
of the weighting function $w(r_i, r_j, R)$ ---
through Eqs.~(\ref{eq:fm}), (\ref{eq:aln}), (\ref{eq:qln}), (\ref{eq:ql}), (\ref{eq:q}), and (\ref{eq:qe}) ---
and their minimization leads to its optimal form for a given or all $R$.

After the weighting function has been determined,
it is time to study the convergence of radial integrals over $r$,
for which the even-tempered scheme of Eq.~(\ref{eq:k})
is natural thanks to its self-similarity.
From Eq.~(\ref{eq:aln}), the sums
\begin{equation}
\label{eq:sr}
\tilde{S}_n (R, t, K) = \sqrt{2}
 \sum\limits_{k=-\infty}^{+\infty} A_{0n} (r_k, R) \, d_k
\end{equation}
add up to make $N$ approximate integrals
\begin{equation}
\bar{S}_n (R, t, K) = \tilde{S}_n (R, t, K) + \tilde{S}_{N+n} (R, t, K) \approx 1
\end{equation}
of the normalized functions of Eq.~(\ref{eq:fa}),
and depend on the point density $K$ and the shift $t$.
The integration error
\begin{equation}
\label{eq:sn}
\EE_n (R, K) = \max_{0\le t\le 1} \left|\bar{S}_n(R, t, K) - 1 \right|
\end{equation}
can be further condensed to
\begin{equation}
\label{eq:sm}
\EE(R, K) = \max_n \EE_n (R, K)
\end{equation}
and even
\begin{equation}
\EE(K) = \max_{R_0 \le R < \infty} \EE(R, K).
\end{equation}
Most often, $\EE(R, K)$ is greatest at $R = R_0$,
and it is enough to work only with $\EE(R_0, K)$,
so we define the function $K(\ee)$ implicitly,
\begin{equation}
\label{eq:se}
\EE\bigl(R_0, K(\ee)\bigr) = \ee ,
\end{equation}
as the (logarithmic) radial point density
needed to reach the integration accuracy $\ee$.

We could have also considered the product $K(\ee)\cdot\EM (\ee)$
as a measure of both radial and angular integration cost
to be minimized, but we put it aside.

\begin{figure}
\includegraphics[scale=1.0]{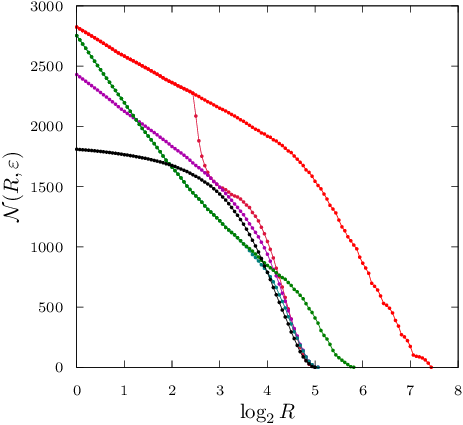}
\caption{\label{fig:n} Functional $\EN(R,\ee)$ of Eq.~(\ref{eq:nre}) computed over: \\
$w_3(x)$ of Eq.~(\ref{eq:wp}) with (red) $\varrho(R) = R$, \\
(crimson) Eq.~(\ref{eq:rcm}), and (purple) Eq.~(\ref{eq:rbc}); \\
$w_4(x)$ of Eq.~(\ref{eq:wp}) with (green) $\varrho(R) = R$ and (teal) Eq.~(\ref{eq:rcm}); \\
(black) $u(r)$ of Eq.~(\ref{eq:ur}) with $\mu=4$, $\nu=8$. \\
$\log_2 \al = -2$ and $\log_2 \ee = -20$ everywhere.}
\end{figure}

We begin our numerical studies with the well-known~\cite{B88} weighting functions of the kind
\begin{equation}
\label{eq:wb}
w(r_i, r_j, R) = w_p \left(\frac{r_i - r_j}{\varrho(R)}\right) ,
\end{equation}
where the simplest smooth step function
\begin{equation}
\label{eq:wp}
w_p(x) =
\left\{ \begin{array}{rcr}
1, && x < -1, \\
\tfrac12 - \tfrac12 s_p(x), && -1\le x \le 1, \\
0, && x > 1,
\end{array}\right.
\end{equation}
is made from the shifted $p$-times iterated polynomial
\begin{eqnarray}
\label{eq:sp}
s_p(x) &=& \tfrac32 s_{p-1}(x) - \tfrac12 s^3_{p-1}(x), \\
s_0(x) &=& x.
\end{eqnarray}
For the distance scale function $\varrho(R)$,
the simplest~\cite{B88} case $\varrho(R) = R$ can be compared to the newer~\cite{LKO18} cut-off version
\begin{equation}
\label{eq:rcm}
\varrho(R) = \min (R, c)
\end{equation}
with some $c$.
For $2^{-4} \le \al \le 2^{-2}$ and $2^{-24} \le \ee \le 2^{-16}$
we have optimized this $c$ and found a fit
\begin{equation}
\label{eq:ca}
c = C/\al
\end{equation}
with $C\approx 1.5$ for $p=3$ and $C\approx 3.0$ for $p=4$, $C$ also being a weak and irregular function of $\ee$.
Fig.~\ref{fig:n} shows a typical example where we see
how Eq.~(\ref{eq:rcm}) helps to overcome the shortcomings of the simplest $\varrho(R) = R$,
strongly for $p=3$, but less so for $p=4$.

For $p=3$, we have also optimized
the values of $\varrho(R)$ for $\EN (R, \ee)$ at each $R_k$ of Eq.~(\ref{eq:rk})
and found them to fit well to a two-parameter function
\begin{equation}
\label{eq:rbc}
\varrho(R) = c \left[1 - \exp\left(-\frac{b R}{c} - \frac{b^2 R^2}{2c^2} \right) \right]
\end{equation}
with $b\approx 0.9$ and $c$ following Eq.~(\ref{eq:ca}) with $C\approx 2.0$ for all $\al$ and $\ee$ studied.
Fig.~\ref{fig:n} shows a further lowering and now a smooth curve,
this seems to be the best one can get from Eq.~(\ref{eq:wb}).
For $p=4$, Eq.~(\ref{eq:rbc}) tends to an unsafe $b>1$, and when constrained to $b=1$,
there is only a slight change from what Eq.~(\ref{eq:rcm}) yields.

Higher derivative discontinuity of $w_p(x)$ at $x=\pm 1$
made us think of fully differentiable analogs,
we have tested
\begin{equation}
\bar{s}_p(x) = \tanh\left(\frac{3^p}{2^p}\cdot\frac{x}{1 - x^2}\right) ,
\end{equation}
which mimics $s_p(x)$ but is not limited to an integer $p$ ---
even with the optimized $p$, however, it worked only slightly worse
than the original $s_p(x)$ of Eq.~(\ref{eq:sp}).

Another kind~\cite{BR66} of weighting function
\begin{equation}
\label{eq:wu}
w(r_i, r_j) = \frac{u(r_i)}{u(r_i) + u(r_j)}
\end{equation}
is made from an unnormalized spherically-symmetric distribution $u(r)\ge 0$,
we tried a number of them until we have found a good and simple analytical form
\begin{equation}
\label{eq:ur}
u(r) = \frac1{r^\nu} \exp\biggl(-\left(\frac{r}{\si}\right)^\mu \biggr)
\end{equation}
with an optimized length scale $\si$.
Seeking the best $\mu$ and $\nu$ among integers,
we have settled on $\mu = 4$ (although $\mu = 6$ worked as well)
and then believed that $\nu = 8$ would have been right also.
Strikingly, as seen in Fig.~\ref{fig:n},
all this yields $\EN(R,\ee)$ that is smooth and everywhere lower
than the best we can get from Eq.~(\ref{eq:wb})!

\begin{figure}
\includegraphics[scale=1.0]{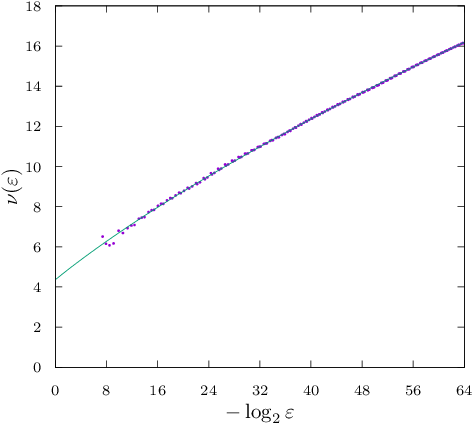}
\caption{\label{fig:nu} Function $\nu(\varepsilon)$ as computed numerically (dots)
 and its approximation (solid line) of Eq.~(\ref{eq:nue}).}
\end{figure}

Further tests have shown, however,
that $\nu$ should be a function of at least $\ee$
lest there be too fast a growth of $\el(r, R, \ee)$
as $\ee\to 0$ at and near $r = R$,
that is when the sphere passes through the other center.
To find our best $\nu(\ee)$,
we minimize $Q_L (1, 1)$ of Eq.~(\ref{eq:ql}) with respect to $\nu$,
using Eqs.~(\ref{eq:wu}) and~(\ref{eq:ur}) with $\si=\infty$,
for $L=8,\dots,192$ and so we get a table of pairs $(\ee_L,\nu_L)$
shown in Fig.~\ref{fig:nu}. A simple function
\begin{equation}
\label{eq:nue}
\nu(\ee) = \kappa\cdot \left(\eta - \log_2 \ee\right)^\zeta
\end{equation}
fits these data well,
the parameters $\kappa\approx 0.748$, $\eta\approx 12.0$, and $\zeta\approx 0.71$
can be determined to only a few digits
because there seems to be a random noise-like component,
but this is enough as seen in Fig.~\ref{fig:nu}.
With $\nu(\ee)$ of Eq.~(\ref{eq:nue}) at hand,
we optimize $\si$ in Eq.~(\ref{eq:ur})
for the ranges $2^{-4}\le\al\le 2^{-2}$ and $2^{-32} \le\ee\le 2^{-16}$,
and we find a good fit to the data with
\begin{equation}
\label{eq:sae}
\si(\al,\ee) = \frac{\sqrt{\beta - \gamma \log_2 \ee }}{\al}
\end{equation}
and the parameters $\beta\approx 1.32$ and $\gamma\approx 0.38$ safely rounded.
Thus we get our best weighting function
of Eqs.~(\ref{eq:ur}), (\ref{eq:nue}), and~(\ref{eq:sae}).

The multicenter~\cite{B88} generalization of Eq.~(\ref{eq:wb})
uses the intermediate products
\begin{equation}
P_i(\vec{r}) = \prod_{j\ne i} w_{ij}(\vec{r}) ,
\end{equation}
the so-called fuzzy Vorono\"i cell functions,
from which the atomic weights are made,
\begin{equation}
W_i(\vec{r}) = \frac{P_i(\vec{r})}{\sum_j P_j(\vec{r})} ,
\end{equation}
without cut-offs, their computational cost grows cubically with the number of atoms.
At the same time, Eq.~(\ref{eq:wu}) readily generalizes into the simplest form,
\begin{equation}
\label{eq:wun}
W_i(\vec{r}) =
\frac{u\big(|\vec{r} - \vec{R}_i |\big)}
     {\sum_j u\big(|\vec{r} - \vec{R}_j |\big)} ,
\end{equation}
growing at most quadratically, making it once again our best scheme.
Using cut-offs, both schemes reach a linear scaling,
but the latter should have a sooner onset and a much smaller prefactor.

Now we need a way to assign the orders $L_i(r)$ of spherical quadrature rules
at distance $r$ around each center $i$ in a polyatomic environment,
they should be no less than
\begin{equation}
L^0_i = 4 l_i + l_\mathrm{d},
\end{equation}
$l_i$ being the highest angular momentum of basis functions on the $i$-th atom,
and $l_\mathrm{d}$ the order of derivatives (if any).
Starting from the two-center distribution $L(r,R,\al,\ee)$,
a rounded up integer version of $\el(r, R, \ee)$ from Eq.~(\ref{eq:qe}),
to which we make a simple analytic fit in Appendix~\ref{sec:fl}
as shown in Fig.~\ref{fig:l},
we try to find a multicenter generalization of diatomic fragment functions
\begin{equation}
L_{ij}(r) = L\bigl(r,\left|\vec{R}_i - \vec{R}_j\right|,\al,\ee \bigr)
\end{equation}
first in the form
\begin{equation}
\label{eq:lij}
\bar{L}_i(r) = \max\bigl(L^0_i, \max_{j\ne i} L_{ij}(r) \bigr) ,
\end{equation}
in other words, $L_{ij}(r)$ is the influence of $j$-th atom
on $L_i(r)$, and the greatest value is taken.
We see in Fig.~\ref{fig:l} the peaks around $r\approx R$,
they should be even sharper for $\bigl(L_i(r) + 1\bigr)^2$,
so it would have been a waste to work with the good-for-all-$R$ solution
\begin{equation}
\tilde{L}_i(r) = \max\bigl(L^0_i, \max_{R>R_0} L(r,R,\al,\ee) \bigr) ,
\end{equation}
and we were optimistic about Eq.~(\ref{eq:lij}) for some time.
Polyatomic tests have later shown, however,
that the influences are not independent and a higher $L_i(r)$
is needed when the other atoms are crowding around.
We find a quick fix to this problem,
\begin{equation}
L_i(r) = \bar{L}_i(r) \cdot \theta\left(\frac{\bar{M}_i(r)}{\bigl(\bar{L}_i(r) + 1\bigr)^2} - 1\right),
\end{equation}
\begin{equation}
\bar{M}_i(r) = (L^0_i+1)^2 + \sum_{j\ne i}\left( \bigl(L_{ij}(r) + 1\bigr)^2 - 1\right),
\end{equation}
\begin{equation}
\theta(x) = 1 + \beta\cdot\bigl(1 - \exp(-\gamma x) \bigr),
\end{equation}
the parameters $\beta\approx \frac12$ and $\gamma\approx 1$ can be adjusted
to get $L_i(r)$ high enough, but often too much.
Thus, a straightforward assignment of quadrature orders
can hardly be as good as we wanted (but we had to have studied it before saying so!),
and we have therefore worked out a new fully adaptive method (given below)
of the old~\cite{AW92} kind.

\begin{figure}
\includegraphics[scale=1.0]{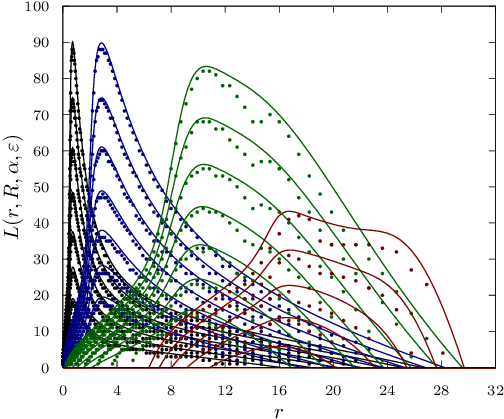}
\caption{\label{fig:l} Distribution of spherical quadrature orders (dots)
and its approximation $L(r,R,\al,\ee)$ of Appendix~\ref{sec:fl}, unrounded (lines),
for $\log_2 \al=-2$;
$-\log_2 \ee =12,16,20,24,28,32,36$;
and $R=1,4,16,32$ (black, blue, green, red).}
\end{figure}

In the end,
we need a better radial integration scheme than in Eq.~(\ref{eq:k}).
We have optimized not just one~\cite{M11}
but the two parameters $p$ and $q$ in the mapping
\begin{equation}
r = \frac1A \exp\bigl(x - q\exp(-px) \bigr)
\end{equation}
of the distance $r$ onto the dimensionless variable $x$
in the range $-\infty < x < \infty$, so that the trapezoidal rule
\begin{equation}
x_k = x_0 + k h
\end{equation}
works well (with cut-offs at both ends)
for a set of functions of Eq.~(\ref{eq:fa}) with $\al_n \le A$,
and for any $x_0$.
A good solution is
$p = 2, q = \tfrac18$,
a detailed derivation is given in Appendix~\ref{sec:rq}
together with the convergence properties as $h\to 0$.
Setting $x_0 = \ln A$, we get the quadrature roots and weights
\begin{equation}
\label{eq:rd}
\left\{
\begin{array}{lcl}
r_k &=& \exp\Bigl(hk - q\exp\bigl(-p(hk + \ln A)\bigr)\Bigr), \\
d_k &=& \Bigl(1 + pq\exp\bigl(-p(hk + \ln A)\bigr)\Bigr) h ,
\end{array}
\right.
\end{equation}
which for $h = (\ln 2)/K$
is a tailored version of the ``half double-exponential''
scheme of Eq.~(\ref{eq:k}),
and a radial integral is computed as
\begin{equation}
\int\limits_0^\infty r^2 f(r)\de r \approx
\sum\limits_{k=k_0}^{k_1} r^3_k f\left(r_k \right)\, d_k .
\end{equation}
The inner cut-off $k_0$ is clearly set by
\begin{equation}
\label{eq:k0}
\frac{A^3}{\pi^{3/2}} r_{k_0}^2 d_{k_0} \approx \ee ,
\end{equation}
(we aim it at the functions $b_n(r)$ of Eq.~(\ref{eq:fa}) and also at $r^{-1} b_n(r)$
as a model of Coulomb integrals, hence $r_{k_0}^2$),
while for the outer,
\begin{equation}
\label{eq:k1}
r_{k_1} \approx r_1(\ee) ,
\end{equation}
we have to study (for our best weighting function)
how far $r_k$ should reach to converge the sum in Eq.~(\ref{eq:sr})
to within $\ee_1 \approx \tfrac14 \ee$, for all $R$,
and we compute a table of values that can be fitted well by the function
\begin{equation}
\label{eq:r1}
r_1(\ee) = \frac1{\al}
\Bigl[\beta + \frac{\kappa}{\gamma}\ln\Bigl(1 + \exp\bigl(\gamma(\eta - \log_2 \ee)\bigr)\Bigr)\Bigr]
\end{equation}
with the parameters
$\beta\approx 4.38$, $\kappa\approx 0.123$,
$\gamma\approx 0.6$, and $\eta\approx -15.0$,
we see that $r_1(\ee)$ stretches beyond the range of functions on the first center
to reach what the weighting function has not fully suppressed on the other.

Solving Eq.~(\ref{eq:se}) with $R_0 =1$
numerically, a table of $K(\ee)$ values for $6\le -\log_2 \ee \le 36$
is computed and can be fitted well by the function
\begin{equation}
\label{eq:ke}
K(\ee) = \beta + \kappa\cdot \left(\eta - \log_2 \ee \right)^\zeta
\end{equation}
with parameters
$\beta\approx 0.91$, $\kappa\approx 0.0608$,
$\eta\approx -4.2$, and $\zeta\approx 1.59$,
the functions on the other center
do make the radial integration more of a challenge
even with the best weighting function we have,
this $K(\ee)$ is up to a few times higher than in the ideal one-center case of Eq~(\ref{eq:e0}).

So we set the range and density of integration points
by Eqs.~(\ref{eq:k0}), (\ref{eq:k1}), (\ref{eq:r1}), and~(\ref{eq:ke}).

Working with the traditional atomic basis functions
\begin{equation}
\label{eq:chi}
\chi_{mlni}(\vec{r}) =
 Y_{lm}\left(\frac{\vec{r} - \vec{R}_i}{|\vec{r} - \vec{R}_i|}\right)
 f_{nli}\bigl(|\vec{r} - \vec{R}_i|\bigr)
\end{equation}
of Gaussian~\cite{B50} type, with the radial parts
\begin{equation}
\label{eq:fchi}
f_{nli}(r) = r^l \sum\limits_p c_{pnli} \exp\left(-a_{pli} r^2\right),
\end{equation}
and the exponent range
\begin{equation}
\label{eq:aminmax}
a^\mathrm{min}_i \le a_{pli} \le a^\mathrm{max}_i,
\end{equation}
we set
\begin{equation}
\label{eq:ai}
A_i = \sqrt{2a^\mathrm{max}_i}
\end{equation}
in Eqs.~(\ref{eq:rd}) and~(\ref{eq:k0}) for each atom $i$,
while a global value
\begin{equation}
\label{eq:a0}
\al = \sqrt{2 \min_i a^\mathrm{min}_i }
\end{equation}
is taken for all atoms in the system.
(We remember that the ``atomic size adjustments''~\cite{B88}
have also been dropped from later works.)

Putting everything together, we get the positions $\{\vec{r}_{mki}\}$ and weights $\{w_{mki}\}$
of the multicenter spatial cubature
\begin{eqnarray}
\label{eq:rmki}
\vec{r}_{mki} &=& \vec{r}_{m L_{ki} ki}, \\
\label{eq:wmki}
w_{mki} &=& w_{m L_{ki} ki}, \\
\vec{r}_{mLki} &=& \vec{R}_i + \vec{X}_i \vec{u}_{mL} r_{ki}, \\
w_{mLki} &=& 4\pi s_{mL} r_{ki}^3 W_i \left(\vec{r}_{mLki}\right) d_k, \\
\label{eq:lki}
L_{ki} &\equiv& L_i (r_{ki}),
\end{eqnarray}
where $\{\vec{u}_{mL}\}$ are the positions and $\{s_{mL}\}$ the weights
of a quadrature~\cite{L76,LL99} of order $L$ on the unit sphere.
For each atom, the spherical grids are aligned
with the principal axes $\vec{X}_i$ of an inertia-like tensor
\begin{equation}
\label{eq:yi}
\vec{Y}_i = \sum\limits_{j\ne i} \vec{Y}\left(\vec{R}_j - \vec{R}_i, \al\right) ,
\end{equation}
\begin{equation}
\label{eq:yra}
\vec{Y}(\vec{r}, \al) = \left(\vec{r}\vec{r}^\mathrm{T} - |\vec{r}|^2 \right) \exp\left(-\al^2 |\vec{r}|^2\right) ,
\end{equation}
to help find a unique orientation.
Any three-dimensional molecular integral can now be evaluated as a simple sum
\begin{equation}
\int f(\vec{r}) \de^3 \vec{r} \approx
\sum\limits_n w_n f\left(\vec{r}_n \right) ,
\end{equation}
where $n = (m,k,i)$ is a combined index.

For a set of atomic basis functions $\{\chi_\mu (\vec{r}) \}$,
their pair product densities
\begin{equation}
q_{\mu\nu}(\vec{r}) = \chi^\dagger_\mu(\vec{r}) \chi_\nu(\vec{r})
\end{equation}
give rise to the overlap integrals
\begin{eqnarray}
S_{\mu\nu} &=& \int q_{\mu\nu}(\vec{r}) \de^3 \vec{r} \\
&\approx& \tilde{S}_{\mu\nu} = \sum\limits_n w_n q_{\mu\nu}\left(\vec{r}_n \right) .
\end{eqnarray}
For the two-electron repulsion integrals
\begin{eqnarray}
V_{\ka\la\mu\nu} &=&
 \int \frac{q_{\ka\la}(\vec{r}_1) q_{\mu\nu}(\vec{r}_2)}{|\vec{r}_1 - \vec{r}_2|}
 \de^3 \vec{r}_1 \de^3 \vec{r}_2 \\
\label{eq:vklmn}
&\approx& \tilde{V}_{\ka\la\mu\nu} =
 \sum\limits_n w_n q_{\ka\la}\left(\vec{r}_n \right) v_{\mu\nu}\left(\vec{r}_n \right) ,
\end{eqnarray}
the seminumerical integration using the analytically computed potentials
\begin{equation}
v_{\mu\nu}(\vec{r}) = \int \frac{q_{\mu\nu}(\vec{r}_2)}{|\vec{r} - \vec{r}_2|} \de^3 \vec{r}_2
\end{equation}
is at the heart of the fast electronic structure methods we want to use.
The integral errors,
\begin{eqnarray}
\label{eq:es}
\ee_S &=& \max_{\mu\nu} \left|\tilde{S}_{\mu\nu} - S_{\mu\nu} \right| , \\
\label{eq:ev}
\ee_V &=& \max_{\ka\la\mu\nu} \left|\tilde{V}_{\ka\la\mu\nu} - V_{\ka\la\mu\nu} \right| ,
\end{eqnarray}
should be small and controllable.
If Eq.~(\ref{eq:vklmn}) were used as written,
the long-range nature of Coulomb interaction would lead to an error
in the molecular energy growing quadratically with its size ---
even though a very fine integration grid may make it small enough,
a smarter way to bring it down to linear
is by replacing~\cite{F87} the densities on the grid $q_{\ka\la}\left(\vec{r}_n \right)$
with their ``corrected'' counterparts
\begin{equation}
\tilde{q}_{\ka\la}(\vec{r}) = \chi_\ka(\vec{r}) \chi_{\la'}(\vec{r}) Z_{\la' \la} ,
\end{equation}
\begin{equation}
\vec{Z} = \tilde{\vec{S}}^{-1} \vec{S} ,
\end{equation}
that yield the exact overlap integrals when summed up over the grid points.
We have found the symmetric correction
\begin{equation}
\bar{q}_{\ka\la}(\vec{r}) = \chi_{\ka'}(\vec{r}) \chi_{\la'}(\vec{r}) O_{\ka' \ka} O_{\la' \la} ,
\end{equation}
\begin{equation}
\vec{O} = \tilde{\vec{S}}^{-1/2} \vec{S}^{1/2} ,
\end{equation}
to work no less well, having a good property
$\bar{q}_{\la\ka}(\vec{r}) = \bar{q}_{\ka\la}(\vec{r})$, unlike $\tilde{q}_{\ka\la}(\vec{r})$,
which may make them easier to handle.
Another way to get rid of the quadratic error growth
is to rearrange the electrostatic terms by adding and subtracting
some promolecule density so that only the Coulomb potential of the deformation density
\begin{equation}
J^\mathrm{d}(\vec{r}) =
\sum\limits_{\mu\nu} v_{\mu\nu}(\vec{r}) D_{\mu\nu} -
\sum\limits_k v_k(\vec{r}) d_k
\end{equation}
is handled by the seminumerical integration
\begin{equation}
\tilde{J}_{\mu\nu}^\mathrm{d} =
 \sum\limits_n w_n q_{\mu\nu}\left(\vec{r}_n \right) J^\mathrm{d}\left(\vec{r}_n \right),
\end{equation}
where $D_{\mu\nu}$ is the molecular density matrix,
and $v_k(\vec{r})$ are potentials of simple (such as Guassian) atom-centered unit-charge distributions
with coefficients $d_k$ on each atom adding up to neutralize the nuclear charge,
we would readily take these from the optimized effective-potential work~\cite{LB20}
to further minimize the errors.

Now, back to the problem of spherical quadrature orders $L_{ki}$
in Eqs.~(\ref{eq:rmki}) and~(\ref{eq:rmki}),
our best solution is an adaptive selection
based on the convergence with $L$ of the surface inegrals
\begin{equation}
\label{eq:slki}
S_{\mu\nu, Lki} = \frac1{d_k} \sum\limits_m w_{mLki}\, q_{\mu\nu}(\vec{r}_{mLki}) v(\vec{r}_{mLki}) ,
\end{equation}
estimated from the differences
\begin{equation}
\label{eq:emnlki}
\EE_{\mu\nu, Lki} = \left| S_{\mu\nu, L + L_1 ,ki} - S_{\mu\nu, Lki} \right| .
\end{equation}
The simplest error measure would have been
\begin{equation}
\EE_{Lki} = \max_{\mu\nu} \EE_{\mu\nu, Lki} ,
\end{equation}
but we want it to be rotationally-invariant,
so we average it over the blocks to get
\begin{equation}
\label{eq:elki}
\bar{\EE}_{Lki} =
\max_{\bar{\mu}\bar{\nu}} \sqrt{
\frac{\sum\limits_{m_{\bar{\mu}} m_{\bar{\nu}}} \EE^2_{m_{\bar{\mu}} m_{\bar{\nu}} ,\bar{\mu}\bar{\nu}, Lki}}
     {2\min(l_{\bar{\mu}}, l_{\bar{\nu}}) + 1}
} ,
\end{equation}
where the combined index $\mu = (m_\mu,l_\mu,n_\mu,i_\mu)$
is split into $m_{\bar{\mu}}$ and
$\bar{\mu} = (l_{\bar{\mu}},n_{\bar{\mu}},i_{\bar{\mu}})$ .
The values of $\bar{\EE}_{Lki}$ are computed for the growing $L$,
mostly in steps of $L_1 = 6$, until
\begin{equation}
\label{eq:e0lki}
\bar{\EE}_{L_{ki} ki} \approx \ee .
\end{equation}
For now, we pick the octahedrally-symmetric spherical grids~\cite{LL99}
in the series of orders
$L=$
    3,  5,  7,  9, 11, 15,  17,  19,  21,  23,  29,  31,  35,  41,  47,  53,   59,   65,   71,   77,   83,   89,   95,  101,  107,  113,  119,  125,  131,
having
    6, 14, 26, 38, 50, 86, 110, 146, 170, 194, 302, 350, 434, 590, 770, 974, 1202, 1454, 1730, 2030, 2354, 2702, 3074, 3470, 3890, 4334, 4802, 5294, 5810 
points, but other choices can be made.

With $v(\vec{r}) \equiv 1$ in Eq.~(\ref{eq:slki}),
we would simply adapt the grids
to an accurate evaluation of the overlap integrals,
but then they might not be as good for the two-electron integrals;
to model the influence of $v_{\mu\nu}\left(\vec{r}_n \right)$
in Eq.~(\ref{eq:vklmn}),
we take
\begin{equation}
\label{eq:v1}
v(\vec{r}) = 1 + \sum\limits_i \frac{\exp\left(-a|\vec{r} - \vec{R}_i|^2 \right)}{|\vec{r} - \vec{R}_i|},
\end{equation}
and find $a=\tfrac14$ to work well.

We have given our method in full and are ready to test it.
To sum it up, it takes as input a molecular geometry $\{\vec{R}_i\}$,
atomic basis functions of Eqs.~(\ref{eq:chi}), (\ref{eq:fchi}), ~(\ref{eq:aminmax}),
and an integration accuracy $\ee$.
The values $A_i$ of Eq.~(\ref{eq:ai}) and $\al$ of Eq.~(\ref{eq:a0})
are used to set up the radial quadrature of Eq.~(\ref{eq:rd})
with $p = 2, q = \tfrac18$, $h = (\ln 2)/K$, and $A\equiv A_i$
for each atom, with the point density $K$ from Eq.~(\ref{eq:ke})
and the ranges from Eqs.~(\ref{eq:k0}), (\ref{eq:k1}), and~(\ref{eq:r1}).
The power $\nu$ of Eq.~(\ref{eq:nue}) and the scale $\si$ of Eq.~(\ref{eq:sae})
are put into the radial functions $u(r)$ of Eq.~(\ref{eq:ur})
from which the normalized atomic weighting functions
$W_i(\vec{r})$ of Eq.~(\ref{eq:wun}) are built.
The $3\times3$ matrices of Eqs.~(\ref{eq:yi}) and~(\ref{eq:yra})
are diagonalized to get the axes $\vec{X}_i$ of the spherical grids
within the multicenter cubature of Eqs.~(\ref{eq:rmki})--(\ref{eq:lki});
the orders $L_{ki}$ are adaptively selected
by testing the convergence of $\bar{\EE}_{Lki}$
from Eqs.~(\ref{eq:elki}), (\ref{eq:slki}), (\ref{eq:emnlki}), and~(\ref{eq:v1}),
until Eq.~(\ref{eq:e0lki}) holds.

\section{Tests}

We have tested our integration method on a set of molecules
made up of light and heavy atoms,
working with an easy-to-use scalar-relativistic approximation~\cite{L19a}
and optimized sets of atomic basis functions~\cite{L19b},
at CCSD~\cite{PB82} geometries,
the results are shown in Table~\ref{tab:mol}.

For the three typical input accuracy levels, $\eta\equiv -\log_2 \ee = 16, 24, 32$,
the output integral errors $\ee_S$ of Eq.~(\ref{eq:es}) and $\ee_V$ of Eq.~(\ref{eq:ev})
are computed over the whole set of overlap and two-electron integrals
and reported alongside the average numbers of grid points per atom.
Ideally, we should have had $\ee_S = \ee_V = \ee$,
but in practice there is some (hopefully small) difference
that characterizes the method.

\begingroup
\squeezetable
\begin{table}
\caption{\label{tab:mol}Molecular tests of integration accuracy.}
\begin{ruledtabular}
\begin{tabular}{llrrrrrrrrr}
 & & \multicolumn{3}{c}{$\eta = 16$} & \multicolumn{3}{c}{$\eta = 24$} & \multicolumn{3}{c}{$\eta = 32$} \\
\cline{3-5}
\cline{6-8}
\cline{9-11}
molecule & basis & $\eta_S$ & $\eta_V$ & $n_\mathrm{a}$ & $\eta_S$ & $\eta_V$ & $n_\mathrm{a}$ & $\eta_S$ & $\eta_V$ & $n_\mathrm{a}$ \\
\hline
H$_2$              & \texttt{L1}  & 16 & 18 & 2042 & 25 & 27 & 10260 & 35 & 36 &      35420 \\
                   & \texttt{L1a} & 16 & 18 & 2184 & 26 & 27 & 10680 & 34 & 37 &      36368 \\
                   & \texttt{L2}  & 15 & 17 & 3124 & 24 & 25 & 12546 & 34 & 36 &      38956 \\
                   & \texttt{L2a} & 15 & 16 & 3194 & 24 & 23 & 13122 & 33 & 35 &      39702 \\
                   & \texttt{L3}  & 14 & 14 & 5314 & 24 & 23 & 16830 & 33 & 32 &      46058 \\
                   & \texttt{L3a} & 14 & 14 & 5422 & 24 & 19 & 17308 & 33 & 27 &      47230 \\
                   & \texttt{L4}  & 13 & 14 & 6756 & 23 & 21 & 19964 & 32 & 28 &      50802 \\
                   & \texttt{L4a} & 13 & 14 & 6816 & 23 & 17 & 20432 & 32 & 25 &      52374 \\
CH$_4$             & \texttt{L1}  & 15 & 17 & 4026 & 22 & 24 & 25269 & 32 & 34 &{\it 102215}\\
                   & \texttt{L2}  & 15 & 17 & 5706 & 21 & 24 & 30470 & 33 & 34 &{\it 112712}\\
C$_2$H$_6$         & \texttt{L1}  & 15 & 17 & 4556 & 22 & 24 & 36246 & 30 & 32 &{\it 143793}\\
C(CH$_3$)$_4$      & \texttt{L1}  & 14 & 17 & 4888 & 22 & 25 & 40903 & 30 & 33 &{\it 168244}\\
H(CC)$_2$H         & \texttt{L1}  & 14 & 16 & 3770 & 23 & 24 & 16772 & 27 & 29 &      53284 \\
H(CC)$_3$H         & \texttt{L1}  & 13 & 15 & 3938 & 23 & 25 & 17442 & 27 & 29 &      56199 \\
H(CC)$_4$H         & \texttt{L1}  & 13 & 15 & 4026 & 23 & 25 & 17654 & 27 & 29 &      57236 \\
H(CC)$_5$H         & \texttt{L1}  & 13 & 15 & 4068 & 22 & 25 & 17732 & 27 & 29 &      57667 \\
Li$_4$F$_4$        & \texttt{L1}  & 16 & 18 & 5545 & 23 & 25 & 36969 & 31 & 33 &{\it 142057}\\
Cs$_4$I$_4$        & \texttt{L1}  & 13 & 17 & 6822 & 23 & 21 & 36827 & 30 & 27 &{\it 143221}\\
Fe(C$_5$H$_5$)$_2$ & \texttt{L1}  & 12 & 14 & 5910 & 21 & 24 & 42898 & 30 & 32 &{\it 170116}\\
UO$_3$             & \texttt{L1}  & 12 & 13 & 7598 & 21 & 21 & 32652 & 28 & 28 &{\it 114061}\\
\end{tabular}
\begin{flushleft}
All $\ee$ values are given as negative binary logarithms $\eta = -\log_2 \ee$: \\
for input values $\eta = 16,24,32$, the observed accuracy \\
of overlap $\ee_S$ and two-electron repulsion $\ee_V$ integrals is listed \\
along with the average number of grid points per atom $n_\mathrm{a}$ \\
(printed in \textit{italics} when running out of spherical grids with $L\le 131$). \\
Molecular geometries are from CCSD~\cite{PB82}/L1~\cite{L19a,L19b}.
\end{flushleft}
\end{ruledtabular}
\end{table}
\endgroup

The example of H$_2$ shows that
the grid point density does depend, but weakly, on the basis set size.
On the polyacetylenes as models of extended systems,
we see how the atom-centered grids saturate with the chain length,
being localized in space as they should be.
Crowded molecules need denser angular grids,
with roughly up to twice as many points.
At the high accuracy end $\ee=2^{-32}$,
our adaptive method may run out of precomputed sperical grids
as it has to stop at $L=131$,
and this is also where the round-off errors
may start to get the upper hand ---
but such high accuracy would hardly be needed,
and the range $2^{-16}\le\ee\le 2^{-24}$ would be enough
for most applications.
High accuracy comes at a high price, but so is the nature of the problem.

\section{Conclusions}

The quality measure we have defined
helps compare, design, and optimize
the weighting functions for multicenter numerical integration in molecules.
In this way, we have found a remarkably simple one of Eq.~(\ref{eq:ur})
working well, as seen from the tests,
and easy to implement with linear system-size scaling,
as well as for periodic systems.
Together with our radial integration scheme
and a few handy fitted functions of accuracy $\ee$ and exponent $\al$,
it makes a black-box numerical tool
for electronic structure calculations.

The seminumerical evaluation of two-electron integrals
may be the shortest path to fast and scalable~\cite{LTKO20} computation
of direct and exchange terms of wavefunction methods ---
we are working toward its use in long-range-corrected~\cite{L19c}
density functional calculations ---
as for the ``pure''~\cite{PBE96} functionals,
we have already upgraded our code~\cite{L97} with the new weighting functions
and are using it to help organic chemists understand reaction mechanisms
in synthesis~\cite{BL20} and catalysis~\cite{VLFSZTZMB20,AO21,APO21}.
We have also implemented the six-dimensional spatial integration
of dispersion-correction density functionals~\cite{VV10}
and it begins being used even to interpret the experimental spectroscopy~\cite{VLZF21}.

Those who hold true to the original~\cite{B88} or modified~\cite{LKO18}
fuzzy Vorono\"i cells may still find our analysis insightful
and enjoy adopting Eq.~(\ref{eq:rbc}) as a smooth and differentiable alternative
to Eq.~(\ref{eq:rcm}).

\section*{Supplementary material}

The code and data for interactive plotting~\cite{gnuplot} of the quadrature order distributions
is available.

\appendix

\section{Analytic fit to distribution of quadrature orders}
\label{sec:fl}

The order of the spherical quadrature $L$ at distance $r$ from the center,
in the presence of the second atom at distance $R$,
for the integration of a set of functions of Eq.~(\ref{eq:fa}) with $\al_n \ge \al$
to an accuracy of $\ee$, is an integer value rounded up from a continuous distribution
\begin{equation}
\label{eq:lr}
L(r, R, \al, \ee) = \lfloor \tfrac12 + \lambda \left(\al r, \al R, -\log_2 \ee \right)\rfloor .
\end{equation}
As a function of $\eta\equiv -\log_2 \ee$, we find
\begin{equation}
\label{eq:fl1}
\lambda (s, t, \eta) = \lambda_3\bigl(\eta,\kappa(s,t),\gamma(s,t),\beta(s,t)\bigr),
\end{equation}
\begin{equation}
\lambda_3 (\eta,\kappa,\gamma,\beta) =
 \max\left(0, \frac\kappa\gamma \Bigl( \exp\bigl(\gamma(\eta - \beta) \bigr) - 1\Bigr)\right),
\end{equation}
to be a rather good fit, even when constrained to $\gamma\ge 0$,
and so we further need the three functions of only two variables
$s\equiv \al r$ and $t\equiv \al R$,
for which we take
\begin{equation}
\kappa(s,t) = \kappa_9\bigl(a_\kappa(t) s,c_0(t),\dots,c_4(t), p_\kappa(t),u_\kappa(t),v_\kappa(t)\bigr),
\end{equation}
\begin{equation}
\begin{array}{l}
\kappa_9(x,c_0,c_1,c_2,c_3,c_4,p,u,v) \\
= \displaystyle
\frac{c_0 + c_1 x + c_2 x^2 + c_3 x^3 + c_4 x^4}
     {\bigl(1 + p(x - u)^2\bigr)\bigl(1 + (x - v)^2 /p \bigr)},
\end{array}
\end{equation}
\begin{equation}
\gamma(s,t) = \gamma_4\left(\frac{s}{s_\gamma(t)} - 1,c_\gamma(t),a_\gamma(t),b_\gamma(t)\right),
\end{equation}
\begin{equation}
\gamma_4 (x,c,a,b) =\frac{c(a+b)}{b\exp(ax) + a\exp(-bx)},
\end{equation}
\begin{equation}
\beta(s,t) = \beta_7 \bigl(s,w_\beta,q_\beta(t),b_0(t),b_1(t),b_2(t),d_\beta(t),a_\beta(t) \bigr),
\end{equation}
\begin{equation}
\begin{array}{l}
\beta_7 (s,w,q,b_0,b_1,b_2,d,a) = \displaystyle
 -w\ln\frac{s}{\sqrt{s^2 + q^2}} \\
\quad + b_0 + b_1 s + (b_2 s)^2 + d\exp(-as),
\end{array}
\end{equation}
and now for the twenty functions of one variable $t$,
we take $w_\beta$ as a constant and
\begin{equation}
c_i(t) = c_{c_i} + \frac{d_{c_i}}{1 + \exp\bigl(-a_{c_i}(t - t_{c_i})\bigr)},
\quad i=0,1,2,
\end{equation}
\begin{equation}
c_i(t) = c_{c_i} + b_{c_i} t + \frac{d_{c_i}}{1 + \exp\bigl(-a_{c_i}(t - t_{c_i})\bigr)},
\quad i=3,4,
\end{equation}
\begin{equation}
p_\kappa(t) = c_{p_\kappa} + \frac{d_{p_\kappa}}{1 + \exp\bigl(-a_{p_\kappa}(t - t_{p_\kappa})\bigr)},
\end{equation}
\begin{equation}
\begin{array}{rcl}
u_\kappa(t) = c_{u_\kappa} &+& \displaystyle
 \frac{d_{1u_\kappa}}{1 + \exp\bigl(-a_{1u_\kappa}(t - t_{1u_\kappa})\bigr)} \\
&+& \displaystyle
 \frac{d_{2u_\kappa}}{1 + \exp\bigl(-a_{2u_\kappa}(t - t_{2u_\kappa})\bigr)},
\end{array}
\end{equation}
\begin{equation}
v_\kappa(t) = c_{v_\kappa} + a_{v_\kappa} t,
\end{equation}
\begin{equation}
a_\kappa(t) = \frac{c_{a_\kappa} + d_{a_\kappa} \exp(-a_{a_\kappa} t)}{t},
\end{equation}
\begin{equation}
c_\gamma(t) = c_{c_\gamma} + \frac{d_{c_\gamma}}{1 + \exp\bigl(a_{c_\gamma}(t - t_{c_\gamma})\bigr)},
\end{equation}
\begin{equation}
a_\gamma(t) = c_{a_\gamma}
 + \frac{d_{a_\gamma}}{a_{a_\gamma}}\ln\Bigl(1 + \exp\bigl(a_{a_\gamma}(t - t_{a_\gamma})\bigr)\Bigr),
\end{equation}
\begin{equation}
b_\gamma(t) = c_{b_\gamma} + d_{b_\gamma} \exp(-a_{b_\gamma}t),
\end{equation}
\begin{equation}
s_\gamma(t) = \frac{a_{s_\gamma} t}{\Bigl(1 + \left(a_{s_\gamma} t/c_{s_\gamma}\right)^8\Bigr)^{1/8}},
\end{equation}
\begin{equation}
b_0(t) = \bar{b}_0(a_{b_0} t),
\end{equation}
\begin{equation}
\bar{b}_0(x) =
\frac{c_{0b_0} + c_{1b_0} x + c_{2b_0} x^2 + c_{3b_0} x^3 + c_{4b_0} x^4}
     {1 + d_{b_0} x + x^2},
\end{equation}
\begin{equation}
\begin{array}{rcl}
b_1(t) &=& -w_{b_1} \ln t + c_{b_1} \\
 &+& \displaystyle
 \frac{d_{b_1} (a_{b_1} + b_{b_1})}
      {b_{b_1}\exp\bigl(a_{b_1}(t - t_{b_1})\bigr) + a_{b_1}\exp\bigl(-b_{b_1}(t - t_{b_1})\bigr)},
\end{array}
\end{equation}
\begin{equation}
b_2(t) = c_{b_2} + \frac{d_{b_2}}{1 + \exp\bigl(a_{b_2}(t - t_{b_2})\bigr)},
\end{equation}
\begin{equation}
d_\beta(t) = c_{d_\beta} + \frac{d_{d_\beta}}{1 + \exp\bigl(-a_{d_\beta}(t - t_{d_\beta})\bigr)},
\end{equation}
\begin{equation}
a_\beta(t) = t_{a_\beta} /t,
\end{equation}
\begin{equation}
\label{eq:fln}
q_\beta(t) = a_{q_\beta} t.
\end{equation}

\begingroup
\squeezetable
\begin{table}
\caption{\label{tab:fl}Parameters of the fitted function $\lambda (s, t, \eta)$ of Eqs.~(\ref{eq:fl1})--(\ref{eq:fln}).}
\begin{ruledtabular}
\begin{tabular}{lD{.}{.}{-1}|lD{.}{.}{-1}}
param. & \multicolumn{1}{c|}{value} &
param. & \multicolumn{1}{c}{value} \\
\hline
$c_{c_0}$       &    0.668676279  & $c_{c_\gamma}$ &    0.0242344273  \\
$d_{c_0}$       &  186.641143     & $d_{c_\gamma}$ &    0.00551760331 \\
$a_{c_0}$       &    0.811747781  & $a_{c_\gamma}$ &    1.46148448    \\
$t_{c_0}$       &    7.91647624   & $t_{c_\gamma}$ &    1.32454270    \\
$c_{c_1}$       &    1.08222907   & $c_{a_\gamma}$ &    0.0677721947  \\
$d_{c_1}$       & -357.497539     & $d_{a_\gamma}$ &   33.0723100     \\
$a_{c_1}$       &    1.04815549   & $a_{a_\gamma}$ &    1.47695611    \\
$t_{c_1}$       &    7.29100798   & $t_{a_\gamma}$ &    8.71490344    \\
$c_{c_2}$       &   -2.38602515   & $c_{b_\gamma}$ &   18.6441594     \\
$d_{c_2}$       &  366.348638     & $d_{b_\gamma}$ &    1.10579589    \\
$a_{c_2}$       &    1.12810997   & $a_{b_\gamma}$ &    0.478453287   \\
$t_{c_2}$       &    7.24506698   & $a_{s_\gamma}$ &    0.667913966   \\
$c_{c_3}$       &    1.36035206   & $c_{s_\gamma}$ &    4.10526088    \\
$b_{c_3}$       &   -0.370781147  & $w_\beta$      &    5.13862547    \\
$d_{c_3}$       & -167.262891     & $c_{0b_0}$     &    0.789609734   \\
$a_{c_3}$       &    1.18103821   & $c_{1b_0}$     &    4.10945429    \\
$t_{c_3}$       &    7.25207491   & $c_{2b_0}$     &  -11.6359735     \\
$c_{c_4}$       &    0.228585146  & $c_{3b_0}$     &   -8.17218131    \\
$b_{c_4}$       &    0.0772105404 & $c_{4b_0}$     &    4.73484405    \\
$d_{c_4}$       &   28.1888933    & $d_{b_0}$      &   -1.66480653    \\
$a_{c_4}$       &    1.22619196   & $a_{b_0}$      &    0.169755366   \\
$t_{c_4}$       &    7.27043411   & $w_{b_1}$      &    0.691117413   \\
$c_{p_\kappa}$  &    0.350950380  & $c_{b_1}$      &   -2.53307029    \\
$d_{p_\kappa}$  &   -0.134561547  & $d_{b_1}$      &   13.3158093     \\
$a_{p_\kappa}$  &    1.71544291   & $a_{b_1}$      &    0.0619037945  \\
$t_{p_\kappa}$  &    2.07928644   & $b_{b_1}$      &    0.929786879   \\
$c_{u_\kappa}$  &    0.424179982  & $t_{b_1}$      &    7.05430131    \\
$d_{1u_\kappa}$ &    2.53098530   & $c_{b_2}$      &   -0.0247806394  \\
$d_{2u_\kappa}$ &   -4.78594889   & $d_{b_2}$      &    1.02341270    \\
$a_{1u_\kappa}$ &    0.947678571  & $a_{b_2}$      &    1.09338554    \\
$a_{2u_\kappa}$ &    0.976692589  & $t_{b_2}$      &    4.34679065    \\
$t_{1u_\kappa}$ &    3.24854968   & $c_{d_\beta}$  &    8.28469547    \\
$t_{2u_\kappa}$ &    7.14486694   & $d_{d_\beta}$  &  128.465198      \\
$c_{v_\kappa}$  &    1.11180008   & $a_{d_\beta}$  &    1.37480180    \\
$a_{v_\kappa}$  &    0.124042954  & $t_{d_\beta}$  &    5.01229765    \\
$c_{a_\kappa}$  &    2.96116919   & $t_{a_\beta}$  &    4.59153158    \\
$d_{a_\kappa}$  &   -0.777060497  & $a_{q_\beta}$  &    0.125754657   \\
$a_{a_\kappa}$  &    0.386868702  &                &                  \\
\end{tabular}
\end{ruledtabular}
\end{table}
\endgroup

All the 75 parameters shown in Table~(\ref{tab:fl})
are optimized for $\log_2 \al = -\tfrac92$ (as an estimate of $\al\to 0$)
on a three-dimensional table of values (given in the supplementary material)
for $\eta = 6,\dots,36$ in steps of 1,
$\{r_k\}$ of Eq.~(\ref{eq:k})
and $\{R_k\}$ of Eq.~(\ref{eq:rk}) both with $K=16$.
We want to err on the safe side, and we minimize
\begin{equation}
U = \sum\limits_{i,j,k} v\Bigl(\bigl(\lambda(s_i,t_j,\eta_k) + 1\bigr)^2 - \bigl(L_{ijk} + 1\bigr)^2 \Bigr)
\end{equation}
where our own error measure function
\begin{equation}
v(x) = x + \frac{\exp(-hx) - 1}h
\end{equation}
is used instead of the least-squares $\bar{v}(x) = x^2$,
it puts a heavier penalty on $x < 0$ (an exponential growth) than on $x > 0$ (close to linear),
we set $h=4$ and get all $\lambda(s_i,t_j,\eta_k) - L_{ijk} > -\tfrac12$
so that the rounded up approximation of Eq.~(\ref{eq:lr}) is never below the exact table value $L_{ijk}$.

To verify the integrity of Eqs.~(\ref{eq:fl1})--(\ref{eq:fln}) and parameters of Table~(\ref{tab:fl}),
and to help implement them,
a \texttt{gnuplot}~\cite{gnuplot} script file for interactive plotting
is included in the supplementary material.

\section{Radial quadrature}
\label{sec:rq}

Here we study the normalized radial functions
\begin{equation}
\pfi_n (r,\al) = c_n \al^{n+1} r^n \exp\left(-\al^2 r^2 \right),
\end{equation}
\begin{equation}
c_0 = \frac{2}{\sqrt{\pi}}, \qquad
c_1 = 2, \qquad
c_n = \frac{2 c_{n-2}}{n - 1},
\end{equation}
with integer $n \ge 0$,
as prototypes of atomic and molecular radial distributions
to be inegrated over $r$ on a grid of points
\begin{equation}
r_k = \rho\bigl((k + t)h \bigr)
\end{equation}
using a coordinate transformation function $\rho(x)$
whose shape can be optimized.
The range of $\al$ is $0 < \al \le A$,
and it can be given as
\begin{equation}
\al = A \exp(a)
\end{equation}
with $a \le 0$.
The sum
\begin{equation}
\label{eq:ssum}
S_n (a, t, h) =
 h \sum\limits_{k=-\infty}^{+\infty} g_n \bigl((t + k)h, a \bigr),
\end{equation}
\begin{equation}
g_n (x, a) = \pfi_n \bigl(\rho(x), A \exp(a) \bigr) \rho'(x),
\end{equation}
approximates the inegral and converges to the exact value
\begin{equation}
\lim_{h\to 0} S_n (a, t, h) = 1,
\end{equation}
it is periodic in the shift $t$,
\begin{equation}
S_n (a, t + 1, h) = S_n (a, t, h),
\end{equation}
and the inegration error can be defined as the worst case
\begin{equation}
\label{eq:eah}
\ep_n(a, h) = \max\limits_{0\le t\le 1} \bigl| S_n (a, t, h) - 1 \bigr|
\end{equation}
for the given $a$ and $h$, and further the overall error is
\begin{equation}
\label{eq:eh}
\ep_n(h) = \max\limits_{a \le 0} \ep_n(a, h) .
\end{equation}
Very soon, as $h\to 0$, only the lowest spectral component
\begin{equation}
\label{eq:s1}
s_n (a, h) = \int\limits_0^1 S_n (a, t, h) \exp(2\pi\im t) \de t
\end{equation}
is needed, thus
\begin{equation}
\ep_n(a, h) \approx \bar{\ep}_n(a, h) = 2 \bigl| s_n (a, h)\bigr| .
\end{equation}
The integral in Eq.~(\ref{eq:s1}) together with the sum in Eq.~(\ref{eq:ssum})
can be unfolded to get
\begin{equation}
\label{eq:si}
s_n (a, h) = \int\limits_{-\infty}^{+\infty}
 g_n (x, a) \exp\left(\frac{2\pi}{h} \im x \right) \de x .
\end{equation}
Now we are ready for work.

First, we take the simplest and most natural function
\begin{equation}
\label{eq:re}
\rho(x) = \frac1A \exp(x),
\end{equation}
so that the integrand
\begin{equation}
g_n (x, a) = c_n \exp\Bigl((n+1)(a+x) - \exp\bigl(2(a+x)\bigr) \Bigr)
\end{equation}
becomes a function only of $a+x$,
thus the error $\ep_n(a, h)$ of Eq.~(\ref{eq:eah})
is the same for all $a$, and we can put $a=0$ and drop it henceforth.
With
\begin{equation}
\om = \frac{2\pi}{h} = \frac{2\pi K}{\ln 2},
\end{equation}
Eq.~(\ref{eq:si}) becomes
\begin{equation}
s_n (\om) = c_n \int\limits_{-\infty}^{+\infty}
 \exp\bigl((n + 1 + \im\om) x - \exp(2x) \bigr) \de x
\end{equation}
and, changing variable from $x$ to $\exp(2x)$,
it can be expressed in terms of the gamma function,
\begin{equation}
s_n (\om) = \frac{c_n}2 \Gamma\left(\frac{n + 1 + \im\om}{2}\right),
\end{equation}
whose well-known asymptotics
\begin{equation}
\lim_{|z|\to\infty} \ln\Gamma(z) =
 \left(z - \tfrac12 \right) \ln z - z + \tfrac12 \ln (2\pi)
\end{equation}
helps us get the long-sought-after answer:
\begin{eqnarray}
\label{eq:e0}
-\log_2 \ep_0 &\approx& \frac{\pi^2 K}{2\ln^2 2} - \tfrac32, \\
\label{eq:e1}
-\log_2 \ep_1 &\approx& \frac{\pi^2 K}{2\ln^2 2}
 - \tfrac12 \log_2 \frac{\pi^2 K}{\ln 2} - \tfrac32, \\
\label{eq:en}
-\log_2 \ep_n &\approx& -\log_2 \ep_{n-2} \nonumber \\
 &-& \log_2 \sqrt{1 + \left[\frac{2\pi K}{(n-1) \ln 2}\right]^2} .
\end{eqnarray}
It is the constant of $\tfrac32$ in Eq.~(\ref{eq:e0})
we saw after having solved Eq.~(\ref{eq:eah}) numerically
and fitting a straight line through the points
that made us believe in the existence of the closed-form expression,
it is remarkable how closely
Eqs.~(\ref{eq:e0}), (\ref{eq:e1}), and~(\ref{eq:en})
fit the exact solutions of Eq.~(\ref{eq:eah}) with Eq.~(\ref{eq:re}) ---
for a given error $\ep$ and a range of $n$,
the almost linear functions $\ep_n(K)$ are easy to invert numerically
to get the grid point density per octave $K$.

Now, we take the function
\begin{equation}
\label{eq:ree}
\rho(x) = \frac1A \exp\bigl(x - q\exp(-px)\bigr)
\end{equation}
that gives a double-exponential decay of the integrand
\begin{eqnarray}
\label{eq:gee}
g_n (x, a, p, q) &=& c_n \exp\bigl[(n + 1)\bigl(a + x - q\exp(-px)\bigr)\bigr]
\nonumber \\
&\times& \exp\Bigl[-\exp\Bigl(2\bigl(a + x - q\exp(-px)\bigr)\Bigr)\Bigr]
\nonumber \\
&\times& \bigl(1 + pq\exp(-px) \bigr)
\end{eqnarray}
at both ends,
\begin{equation}
\label{eq:l+}
\lim_{x\to +\infty} g_n =
c_n \exp\bigl[(n + 1)(a + x) - \exp\bigl(2(a + x)\bigr)\bigr] ,
\end{equation}
\begin{equation}
\label{eq:l-}
\lim_{x\to -\infty} g_n =
c_n pq\exp\bigl[-px - (n + 1)q\exp(-px)\bigr] ,
\end{equation}
and we want to optimize its parameters $p, q > 0$.
Whenever $q > 0$, the error $\ep_n(a, h, p, q)$ of Eq.~(\ref{eq:eah})
is a nonconstant function of $a$ whose values for some $a\approx 0$
are greater than when $q = 0$, the above case of Eq.~(\ref{eq:re}),
but nearly the same as $a\to -\infty$,
so we have to sacrifice some accuracy at one end ---
however arbitrary it may be, we set the error of Eq.~(\ref{eq:eh}) as
\begin{equation}
\label{eq:e2}
\ep_n\bigl(h,p,q_n(h,p)\bigr) = 2\ep_n(h,0,0)
\end{equation}
and thus get an implicit equation for $q_n(h,p) > 0$,
what is left is to find a good $p$.
(With Eq.~(\ref{eq:gee}), there seem to be no closed-form solutions
for Eqs.~(\ref{eq:ssum}), (\ref{eq:eah}), (\ref{eq:eh}), and even~(\ref{eq:s1}) ---
we have to do it all numerically.)

One way to pin down the value of $p$ is by asking for an equal decay rate
in Eqs.~(\ref{eq:l+}) and~(\ref{eq:l-}), so we get $p=2$.
Solving Eq.~(\ref{eq:e2}) numerically,
we get the values of $q_n(h,2)$ that oscillate (much for $n=0,1$ and less and less for $n>1$)
but seem to have a limit as $h\to 0$.
The case of $n=0$ stands out as the maximum in Eq.~(\ref{eq:eh})
is at $a<0$, all $n \ge 1$ seem to have it at $a=0$.
As a rule, we see $q_{n+1}(h,p) > q_n(h,p)$ and they seem to converge with $n$.
To get one $q$ for a given $p$ and all $h$ and $n$, we need to bracket $q_n(h,p)$ from below,
thus we get a pair
\begin{equation}
\label{eq:p2}
p=2, \quad q\approx\tfrac18,
\end{equation}
for $n\ge0$, and a narrower $q\approx\tfrac13$ for $n\ge1$.

For a full optimization of $p$ and $q$,
we need an objective function,
and we define one such by the implicit equation
\begin{equation}
g_n\bigl(z_n(h,p),0,p,q_n(h,p)\bigr) = \ep_n(h,0,0) 
\end{equation}
and maximize $z_n(h,p)$ over $p$ for a given $h$ to get $p_n(h)$
that yields the fastest decay of the integrand of Eq.~(\ref{eq:gee})
down to $\ep$ as $x\to -\infty$.
The values of $p$ so calculated oscillate around $p\approx 2$
as $h\to 0$ for $n=0$, and slowly grow with $n>0$
--- this only confirms the goodness of Eq.~(\ref{eq:p2})
we now take as our best solution,
and the soundness of arguments behind it.


%
\end{document}